\documentclass[pre,aps,showpacs,floatfix]{revtex4}
\usepackage{graphicx,amsfonts}
\usepackage{epsf, subfigure, verbatim, epsfig}
\newlength{\figwidth}
\begin{document}
%\draft

%version 11 June 2004
\title {\Large \bf  A method to discern complexity in two-dimensional patterns \\
generated by coupled map lattices}

\author{Juan S\'anchez$^1$ and  Ricardo L\'opez-Ruiz$^{2}$} 
 \affiliation{$^1$
 Facultad de Ingenier\'{\i}a, Universidad nacional de Mar del Plata, 
  Av. J.B. Justo 4302, Mar del Plata 7600, Argentina} 
\affiliation{$^2$
DIIS and BIFI, Facultad de Ciencias, 
Universidad de Zaragoza, 50009 - Zaragoza, Spain.}

\begin{abstract}
Complex patterns generated by the time evolution of a one-dimensional digitalized coupled map 
lattice are quantitatively analyzed. A method for discerning complexity among the different 
patterns is implemented. The quantitative results indicate two zones in parameter space
where the dynamics shows the most complex patterns. These zones are located on the two edges 
of an absorbent region where the system displays spatio-temporal intermittency.     

\end{abstract}


\maketitle

%\section{}

This century has been told to be the century of {\it complexity} \cite{hawking}.
Nowadays the question {\it`what is complexity?'} is circulating over 
the scientific crossroads of physics, biology, mathematics and computer science, 
although under the present understanding of the world there seems to be no urgency
to answer the above question. However, many different points of view have been developed 
to this respect and hence a lot of different
answers can be found, at present, in the literature \cite{perakh}.
 
It should be kept in mind that in ancient epochs,
time, space, mass, velocity, charge, color, etc. were only perceptions.
Only after that they became concepts, different tools and instruments were invented for 
quantifying those perceptions, and, finally, only with numbers the scientific laws emerge. 
In this sense, if by complexity it is to be understood that property present in all systems 
attached under the epigraph of `complex systems', this property should be reasonably quantified 
by the different measures that were proposed in the last years.
This kind of indicators is found in those fields where the concept of information
is crucial. Thus, the effective measure of complexity
\cite{grassberger}, the thermodynamical depth \cite{lloyd} and the
simple measure of complexity \cite{shiner} come from physics 
and  other attempts  such as algorithmic complexity \cite{kolmogorov, chaitin}, 
Lempel-Ziv complexity \cite{lempel} and $\epsilon$-machine complexity
\cite{crutchfield} arise from the field of computational sciences.

In particular, taking into account the statistical properties of a system,
an indicator called the {\it LMC (L\'{o}pezRuiz-Mancini-Calbet) complexity}
has been introduced  \cite{lopezruiz95}. 
This magnitude identifies the entropy or information stored in a system and its disequilibrium
i.e., the distance from its actual state to the 
probability distribution of equilibrium, as the two basic ingredients for calculating
its complexity. If $H$ denotes the {\it information} stored in the system and $D$ is its 
{\it disequilibrium},
the LMC complexity $C$ is given by the formula:
\begin{eqnarray}
& C(\{p_i\})  =  H(\{p_i\})\cdot D(\{p_i\})\; = & \nonumber \\
   & =-k \left( \sum_{i=1}^N p_i\log p_i \right)\cdot
    \left( \sum_{i=1}^N \,\left( p_i - \frac{1}{N} \right)^2\right)\, &
    \label{eq:def-c}
\end{eqnarray}
where $\{p_i\}$, with $p_i\geq 0$ and $i=1,\cdots,N$, represents the
distribution of the $N$ accessible states to the system, and $k$ is a constant. 

As it can be straightforwardly seen, the LMC complexity vanishes 
both for completely ordered and for completely 
random systems as it is required for the correct asymptotic properties 
of a such well-behaved measure. Recently, it has been successfully used to
discern situations regarded as complex in discrete systems out of
equilibrium \cite{plastino,calbet,martin,guozhang,zuguo,rosso}. 

As an example, the local transition to chaos via
intermittency \cite{pomeau80} in the logistic map, $x_{n+1}=\lambda x_n(1-x_n)$
presents a sharp transition when C is plotted
versus the parameter $\lambda$ in the region around
the instability for $\lambda\sim \lambda_t=3.8284$. 
When $\lambda<\lambda_t$ the system approaches the laminar regime and 
the bursts become more unpredictable. The complexity increases. When the point
$\lambda=\lambda_t$ is reached a drop to zero occurs for the magnitude $C$.
The system is now periodic and it has lost its complexity.
The dynamical behavior of the system is finally well reflected in the magnitude $C$
(see Fig. 3a of Ref. \cite{lopezruiz95}).

When a one-dimensional array of such maps is put together a more complex behavior
can be obtained depending on the coupling among the units. Ergo the phenomenon
called {\it spatial-temporal intermittency} can emerge \cite{chate87,chate88,jensen90,jensen98}. 
This dynamical regime corresponds 
with a situation where each unit is weakly oscillating around a laminar state
that is aperiodically and strongly perturbed for a traveling burst. 
In this case, the plot of the one-dimensional lattice evolving in time 
gives rise to complex patterns on the plane. If the coupling among units
is modified the system can settle down in an absorbing phase where its dynamics 
is trivial \cite{coullet,toral} and then homogeneous patterns are obtained.
Therefore an abrupt transition to spatio-temporal intermittency can be depicted by
the system \cite{pomeau86,sinha} when modifying the coupling parameter.   

In this letter we are concerned with measuring $C$ in a such
transition for a coupled map lattice of logistic type.
Our system will be a line of sites, $i=1,\ldots,L$, with periodic
boundary conditions. In each site $i$ a local variable $x_i^{n}$ evolves 
in time ($n$) according to a discrete logistic equation. The interaction 
with the nearest neighbors takes place via a multiplicative coupling:
\begin{equation}
x_i^{n+1} = (4-3pX_i^{n})x_i^{n}(1-x_i^{n}).
\label{eq:xn}
\end{equation}  
The variable $X_i^{n}$ is the digitalized local mean field,
\begin{equation}
X_i^{n} = nint \left[\frac{1}{2}\: ({x_{i+1}^{n}+x_{i-1}^{n}}) \right] \: ,
\end{equation}
with {\it $nint(.)$} the integer function rounding its argument to the nearest integer,
and $p$ is the parameter of the system measuring the strength of the coupling ($0<p<1$).  

\begin{figure}[h]
  \centering
  {\includegraphics[angle=0, width=15cm]{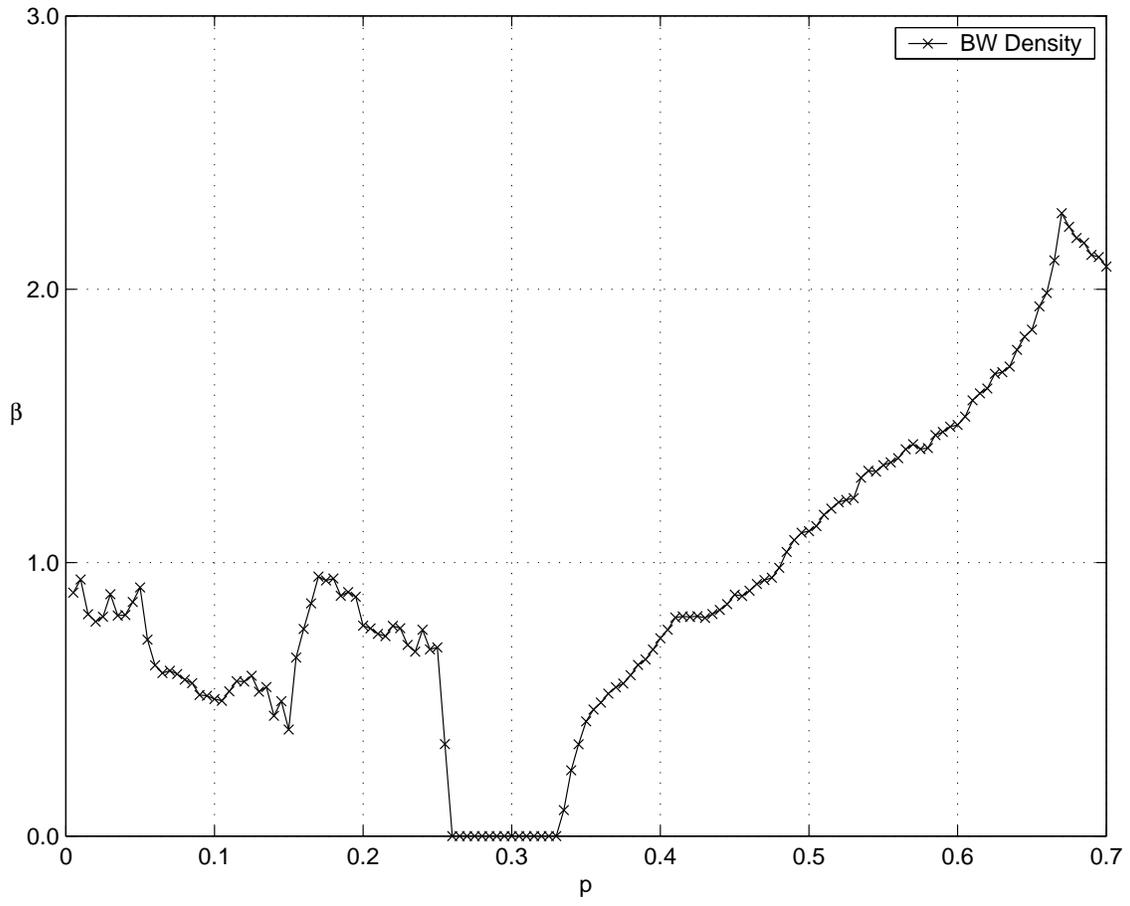}}
  \caption{($\times$) $\beta$ versus $p$. The $\beta$-statistics (or BW density) 
  for each $p$ is the rate between the number of {\it black} and {\it white} cells 
  depicted by the system in the two-dimensional representation of its after-transient time evolution.
  (Computations have been performed with $\Delta p=0.01$ for a lattice of $10000$ sites after a transient 
  of $5000$ iterations and a running of other $5000$ iterations).}
  \label{fig1}
\end{figure}

There is a biological motivation behind this kind of systems \cite{lopezruiz04,lopezruiz05}.
It could represent a {\it colony of interacting competitive individuals}.
They evolve randomly when they are independent ($p=0$). If some competitive interaction 
($p>0$) among them takes place the local dynamics loses its erratic component and becomes
chaotic or periodic in time depending on how populated the vicinity is.
Hence, for bigger $X_i^n$ more populated is the neighborhood of the individual $i$ and 
more constrained is its free action. At a first sight, it would seem that some particular 
values of $p$ could stabilize the system. In fact, this is the case. 
Let us choose a number of individuals for the colony ($L=500$ for instance),
let us initialize it randomly in the range $0<x_i<1$  and 
let it evolve until the stationary regime is attained.
Then the {\it black/white} statistics of the system is performed. That is,
the state of the variable $x_i$ is compared with the critical level $0.5$ for $i=1,\ldots,L$:
if $x_i>0.5$ the site $i$ is considered {\it white} (high density cell) and a counter $N_w$ is 
increased by one, or if $x_i<0.5$ the site $i$ is considered {\it black} (low density cell) and 
a counter $N_b$ is increased by one. This process is executed in the stationary regime
for a set of iterations. The {\it black/white} statistics is then the rate $\beta=N_b/N_w$.
If $\beta$ is plotted versus the coupling parameter $p$ the Figure \ref{fig1} is obtained.

\begin{figure}[h]
  \centering
  {\includegraphics[angle=0, width=15cm]{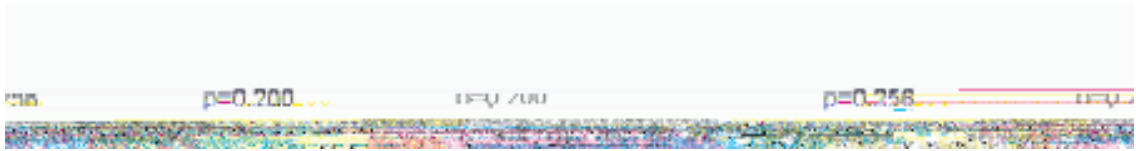}}
  \caption{Digitalized plot of the one-dimensional coupled map lattice (axe OX) evolving in time (axe OY)
  according to Eq. (\ref{eq:xn}): if $x_i^n>0.5$ the $(i,n)$-cell is put in white color 
  and if $x_i^n<0.5$ the $(i,n)$-cell is put in black color. The discrete time
  $n$ is reset to zero after the transitory. (Lattices of $300\times 300$ sites, 
  i.e., $0<i<300$ and $0<n<300$).}
  \label{fig2}
\end{figure}

The region $0.258<p<0.335$ where $\beta$ vanishes is remarkable. 
As stated above, $\beta$ represents the rate between the number of black cells and the number of white 
cells appearing in the two-dimensional digitalized representation of the colony evolution.
A whole white pattern is obtained for this range
of $p$. The phenomenon of spatio-temporal intermittency is displayed by the system 
in the two borders of this parameter region (Fig. \ref{fig2}). 
Bursts of low density (black color) travel in an
irregular way through the high density regions (white color). 
In this case two-dimensional complex patterns 
are shown by the time evolution of the system (Fig.  \ref{fig2}b-c). 
If the coupling $p$ is far enough from this region, i.e., $p<0.25$ or $p>0.4$,
the absorbent region loses its influence on the global dynamics and less structured 
and more random patterns than before are obtained (Fig.  \ref{fig2}a-d).

\begin{figure}[h]
  \centering
  {\includegraphics[angle=0, width=15cm]{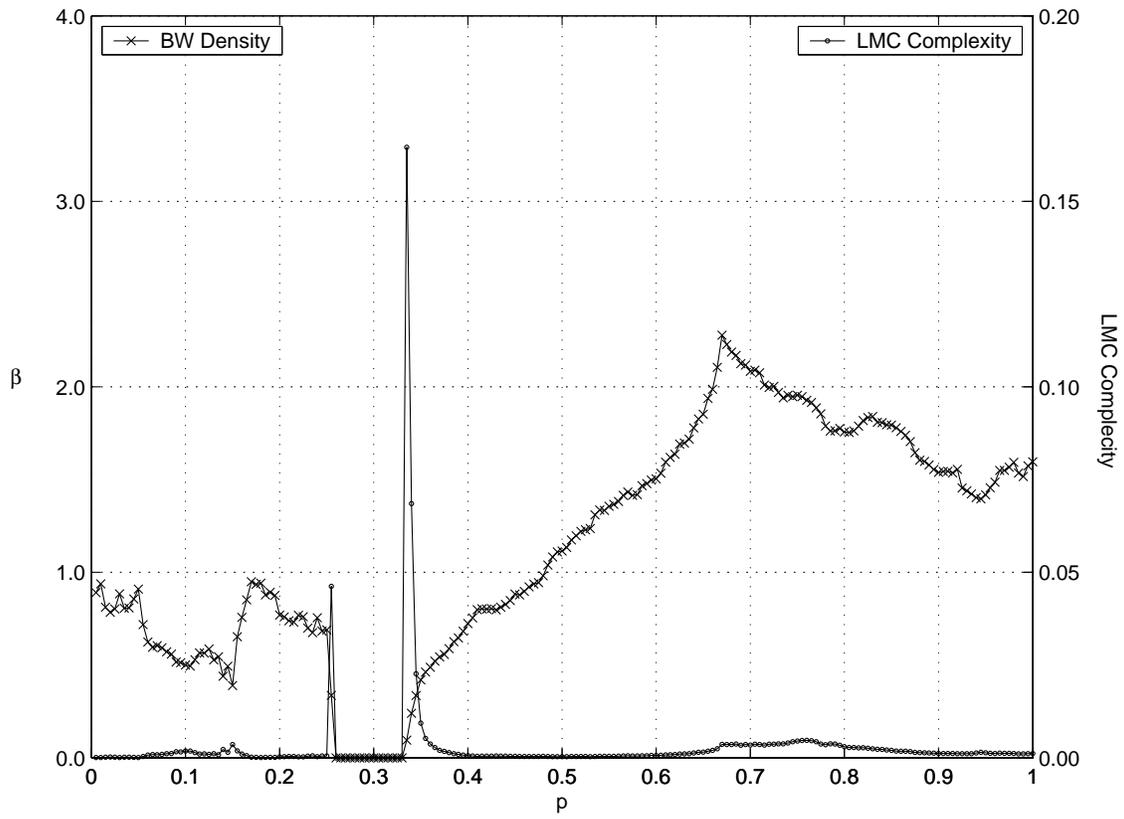}}
  \caption{($\Box$) $C$ versus $p$. Observe the peaks of the LMC complexity located
  just on the borders of the absorbent region $0.258<p<0.335$,
  where $\beta=0$ ($\times$). 
  (Computations have been performed with $\Delta p=0.01$ for a lattice of $10000$ sites 
  after a transient of $5000$ iterations and a running of other $5000$ iterations).}
  \label{fig3}
\end{figure}

If the LMC complexity is quantified as function of $p$, 
our {\it intuition} is confirmed.
The method proposed in Ref. \cite{lopezruiz95} to calculate $C$ is now adapted
to the case of two-dimensional patterns. We let the system evolve
until the stationary regime is attained. 
The variable $x_i^n$ in each cell is successively transformed in a binary sequence
($0$ if $x_i^n<0.5$ and $1$ if $x_i^n>0.5$) when  $i$ covers the lattice, 
$i=1,\ldots,L$, and when $n$ is consecutively increased.
This binary string is analyzed in blocks of $n_o$ bits, where
$n_o$ can be considered the scale of observation.
The accessible states to the system among the $2^{n_o}$ possible states 
is found as well as their probabilities.
Then, the magnitudes $H$, $D$ and $C$ are directly calculated.   
Figure \ref{fig2} shows the result for the case of $n_o=10$. 
Let us observe that the highest $C$ is reached when the dynamics displays 
spatio-temporal intermittency, that is, the {\it most complex patterns} 
are obtained for those values of $p$ that are located
on the borders of the absorbent region $0.258<p<0.335$.
Thus the plot of $C$ versus $p$ shows two tight peaks around the values
$p=0.256$ and $p=0.34$ (Fig. \ref{fig3}).

If the detection of complexity in the two-dimensional case requires to identify 
some sharp change when comparing different patterns, 
those regions in the parameter space
where an abrupt transition happens should be explored 
in order to obtain the most complex patterns.
Smoothness seems not to be at the origin 
of complexity.  As well as a selected few distinct molecules 
among all the possible are in the basis of life \cite{mckay},
discreteness and its spiky appearance 
could indicate the way towards complexity.
Let us recall that the distributions 
with the highest LMC complexity are just those distributions  
with a spiky-like appearance \cite{calbet}. 
In this line, the striking result here exposed confirms the capability  of the 
LMC complexity for signaling a transition to complex behavior when regarding 
two-dimensional patterns.

\newpage
\begin{center} {\bf Acknowledgments} \end{center}
The authors would like to express their gratitude to Internet.
It has permitted them to achieve the present transatlantic collaboration.
They hope to have the possibility to shake their hands some day in the next future.

\newpage

\end{document}